\documentclass[letter]{aa} 

\usepackage{pgf}
\usepackage{graphicx}
\usepackage{txfonts}
\usepackage{natbib,twoopt}
\usepackage[breaklinks=true]{hyperref} 
\bibpunct{(}{)}{;}{a}{}{,} 

\usepackage[utf8]{inputenc}
\usepackage{graphicx}
\usepackage{txfonts}
\newcommand{\kms}{\mbox{$\rm{km}\,s^{-1}$}}

\newcommand{\halpha}{\,H$\alpha$}

\begin{document}

   \title{The transitional gap transient AT~2018hso: new insights on the luminous red nova phenomenon}

\author{Y-Z. Cai \inst{\ref{inst1},\ref{inst2}\thanks{E-mail: yongzhi.cai@studenti.unipd.it}}
\and A.~Pastorello \inst{\ref{inst2}}
\and M. Fraser \inst{\ref{inst3}}
\and S.~J. Prentice \inst{\ref{inst4}}
\and T. M. Reynolds \inst{\ref{inst5}}
\and E. Cappellaro \inst{\ref{inst2}}
\and S. Benetti \inst{\ref{inst2}} 
\and A. Morales-Garoffolo \inst{\ref{inst6}}
\and A.~Reguitti \inst{\ref{inst7},\ref{inst8},\ref{inst1}}
\and N.~Elias-Rosa \inst{\ref{inst2}}
\and S. Brennan  \inst{\ref{inst3}}
\and E.~Callis \inst{\ref{inst3}}
\and G. Cannizzaro  \inst{\ref{inst9},\ref{inst10}}
\and A. Fiore \inst{\ref{inst1},\ref{inst2}}
\and M.~Gromadzki \inst{\ref{inst11}}
\and F.~J.~Galindo-Guil \inst{\ref{inst12},\ref{inst13}}
\and C. Gall \inst{\ref{inst14}}
\and T. Heikkil\"a  \inst{\ref{inst5}}
\and E.~Mason \inst{\ref{inst15}}
\and S.~Moran  \inst{\ref{inst5},\ref{inst12}}
\and F. Onori \inst{\ref{inst16}}
\and A.~Sagu\'es Carracedo \inst{\ref{inst17}}
\and G.~Valerin  \inst{\ref{inst1},\ref{inst2}}
}

\institute{
Universit\`a degli Studi di Padova, Dipartimento di Fisica e Astronomia, Vicolo dell'Osservatorio 2, 35122 Padova, Italy\label{inst1}
\and INAF - Osservatorio Astronomico di Padova, Vicolo dell'Osservatorio 5, 35122 Padova, Italy\label{inst2}
\and School of Physics, O'Brien Centre for Science North, University College Dublin, Belfield, Dublin 4, Ireland \label{inst3}
\and School of Physics, Trinity College Dublin, The University of Dublin, Dublin 2, Ireland \label{inst4}
\and Department of Physics and Astronomy, University of Turku, FI-20014 Turku, Finland \label{inst5}
\and Department of Applied Physics, University of C\'adiz, Campus of Puerto Real, E-11510 C\'adiz, Spain\label{inst6}
\and Millennium Institute of Astrophysics (MAS), Nuncio Monsenor S\`{o}tero Sanz 100, Providencia, Santiago, Chile \label{inst7}
\and Departamento de Ciencias Fisicas, Universidad Andres Bello, Fernandez Concha 700, Las Condes, Santiago, Chile \label{inst8}
\and SRON, Netherlands Institute for Space Research, Sorbonnelaan, 2, NL-3584CA Utrecht, Netherlands \label{inst9}
\and Department of Astrophysics/IMAPP, Radboud University, P.O. Box 9010, 6500 GL Nijmegen, Netherlands \label{inst10}
\and Astronomical Observatory, University of Warsaw, Al. Ujazdowskie 4, 00-478 Warszawa, Poland \label{inst11}
\and Nordic Optical Telescope, Apartado 474, E-38700 Santa Cruz de La Palma, Spain \label{inst12}
\and Depto. de  Astrof\'isica, Centro de Astrobiolog\'ia  (INTA-CSIC), Camino  Bajo  del  Castillo  s/n.  28692, Madrid, Spain \label{inst13}
\and Niels Bohr Institute, University of Copenhagen, Juliane Maries Vej 30, DK-2100 Copenhagen, Denmark \label{inst14}
\and INAF - Osservatorio Astronomico di Trieste, Via Giambattista Tiepolo 11, 34143, Trieste, Italy\label{inst15}
\and Istituto di Astrofisica e Planetologia Spaziali (INAF), via del Fosso del Cavaliere 100, Roma, I-00133, Italy \label{inst16}
\and The Oskar Klein Centre, Physics Department, Stockholm University, SE 106 91 Stockholm, Sweden  \label{inst17}
}

   \date{Received Month Day, 2019; accepted Month Day, 2019}

 \abstract
 {Luminous red novae (LRNe) have absolute magnitudes intermediate between novae and supernovae, and show a relatively homogeneous spectro-photometric evolution. Although they were thought to derive from core instabilities in single stars, there is growing support to the idea that they are triggered by binary interaction, possibly ending with the merging of the two stars. }
{AT~2018hso is a new transient showing transitional properties between those of LRNe and the class of intermediate luminosity red transients (ILRTs) similar to SN~2008S. Through the detailed analysis of the observed parameters, our study support that it actually belongs to the LRN class, and was likely produced by the coalescence of two massive stars.}
{We obtained ten months of optical and near infrared photometric monitoring, and eleven epochs of low-resolution optical spectroscopy of AT~2018hso. We compared its observed properties with those of other ILRTs and LRNe. We also inspected archive Hubble Space Telescope (HST) images obtained about 15 years ago to constrain the progenitor's properties. }
{The light curves of AT~2018hso show a first sharp peak ($M_r = -13.93$ mag), followed by a  broader and shallower second peak, that resembles a plateau in the optical bands. The spectra  dramatically change with time. Early time spectra show prominent Balmer emission lines and a  weak [Ca~{\sc ii}] doublet, which is usually observed in ILRTs. However, the major decrease in the continuum temperature, the appearance of narrow metal absorption lines, the major change in the H$\alpha$ strength and profile, and the emergence of molecular bands support a LRN classification. The possible detection of an $I \sim -8$ mag source at the position of AT~2018hso in HST archive images is consistent with  expectations for a pre-merger massive binary, similar to the precursor of the 2015 LRN in M101.}
{We provide reasonable arguments to support a LRN classification for AT~2018hso. This study reveals growing heterogeneity in the observables of LRNe than thought in the past, making sometimes tricky the discrimination between LRNe and ILRTs. This suggests the need of monitoring the entire evolution of gap transients to avoid misclassifications.}

\authorrunning{Y-Z. Cai et al.}
\titlerunning{The transitional gap transient:  AT~2018hso}

 \maketitle
%

\section{Introduction}\label{Introduction}
Modern all-sky surveys are discovering stellar transients with intrinsic luminosities lying in the middle between those of core-collapse supernovae (SN) and  classical novae (-15 $\leqslant$ $M_\mathrm{V}$ $\leqslant$ -10 mag). These are collectively known as "gap transients" \citep[e.g., ][]{Kasliwal2012PASA...29..482K, Pastorello2019NatAs...3..676P}, and include intermediate luminosity red transients \citep[ILRTs; e.g.,][]{Botticella2009MNRAS.398.1041B, Thompson2009ApJ...705.1364T, Berger2009ApJ...699.1850B}, and luminous red novae \citep[LRNe; e.g.,][]{Munari2002A&A...389L..51M,Tylenda2005A&A...436.1009T,Tylenda2011A&A...528A.114T,Williams2015ApJ...805L..18W,Goranskij2016AstBu..71...82G,Lipunov2017MNRAS.470.2339L}.     

ILRTs have single-peaked light curves resembling those of faint SNe IIP or IIL. Their spectra show prominent Balmer emission features, along with weak Fe {\sc ii}, Na {\sc i} D and Ca {\sc ii} lines. In particular, the [Ca {\sc ii}] doublet feature, prominent at all phases, is a typical signature of ILRTs. These transients are physically consistent with electron-capture induced supernova (EC SN) explosions from super-asymptotic giant branch stars \citep[S-AGB; see, e.g., ][]{Pumo2009ApJ...705L.138P}. Well-studied objects are SN 2008S \citep[][]{Prieto2008ApJ...681L...9P, Botticella2009MNRAS.398.1041B, Smith2009ApJ...697L..49S, Adams2016MNRAS.460.1645A}, NGC 300 OT2008-1 \citep[][]{ Bond2009ApJ...695L.154B, Berger2009ApJ...699.1850B, Humphreys2011ApJ...743..118H}, and AT 2017be \citep{Cai2018MNRAS.480.3424C}.    

LRNe make a distinct group of gap transients, which usually display double-peaked light curves \citep[e.g.,][]{Kankare2015A&A...581L...4K, Blagorodnova2017ApJ...834..107B} and a peculiar spectral evolution with time. In particular, a forest of narrow metal lines in absorption are detected in the spectra during the second light curve maximum, and broad molecular absorption bands (e.g., CaH, CN, TiO, and VO) are observed at very late epochs \citep{Mason2010A&A...516A.108M, Barsukova2014AstBu..69...67B, Blagorodnova2017ApJ...834..107B}. LRNe are very likely produced by extreme  interaction in a close binary system leading to common-envelope (CE) ejection. The final outcome is likely a merger \citep[e.g.,][]{Tylenda2011A&A...528A.114T, Kochanek2014MNRAS.443.1319K, Pejcha2016MNRAS.461.2527P, Pejcha2017ApJ...850...59P, Metzger2017MNRAS.471.3200M, MacLeod2017ApJ...835..282M}. 

While most LRNe have a robust classification, occasionally their discrimination from ILRTs is a difficult task \citep[see, e.g., the controversial cases of M85-2006OT1 and PTF~10fqs; ][]{Kulkarni2007Natur.447..458K, Pastorello2007Natur.449E...1P, Kasliwal2011ApJ...730..134K}. In this context, we report the study of  AT 2018hso, which shows hybrid spectro-photometric properties that challenge the existing paradigm of LRNe.     
  

\section{Object information}
AT 2018hso (also known as ZTF18acbwfza) was discovered on 2018 October 31.53 (UT) by the Zwicky Transient Facility (ZTF\footnote{ZTF: \url{https://www.ztf.caltech.edu/}})  \citep{De2018ATel12162....1D}. Its coordinates are RA=$11^{h}33^{m}51.96^{s}$, Dec=$+53\degr07\arcmin07.10\arcsec$~[J2000], $24.9\arcsec$ south, $23.7\arcsec$ east of the core of the face-on, late-type galaxy NGC~3729. 

The distances to NGC~3729 reported in the NASA/IPAC Extragalactic Database (NED\footnote{NED: \url{http://nedwww.ipac.caltech.edu/}}) are based on the Tully-Fisher method \citep[e.g.,][]{Willick1997ApJS..109..333W, Tully2009AJ....138..323T}, and range from 21.10 to 21.88 Mpc. The kinematic distance corrected for Virgo Infall and obtained adopting a standard cosmology ($H_{0}$=73 \kms $\rm{Mpc}^{-1}$, $\Omega_{M}$=0.27, $\Omega_{\Lambda}$=0.73)  is $d_{k}=20.80 \pm 1.5$ Mpc \citep[][]{Mould2000ApJ...545..547M}, and is in good agreement with Tully-Fisher estimates. Hence, hereafter, we will adopt as distance to NGC~3729 the weighted average of the above values, $d=21.26 \pm 0.56$ Mpc, which provides a distance modulus $\mu = 31.64 \pm 0.06$~$\mathrm{mag}$.   

The Galactic reddening at the coordinates of AT~2018hso is very small, $E(B-V)_{\mathrm{Gal}}$ = 0.01~$\mathrm{mag}$ \citep{Schlafly2011ApJ...737..103S}. Early spectra have modest resolution and limited signal-to-noise (SNR), while late-time spectra are affected by the presence of a forest of absorption metal lines. Yet, for an indicative estimate of the host galaxy reddening, we average the measured equivalent widths (EWs) of the Na {\sc i} D absorption in two early spectra (at $-3.8$ and $+11.3$ d),  obtaining EW$=1.8 \pm 0.5 \AA$. Following the relation in \citet{Turatto2003fthp.conf..200T}, we obtain a host galaxy reddening $E(B-V)_{\mathrm{Host}}=0.29$~$\pm$ 0.08 $\mathrm{mag}$. The total line-of-sight reddening is therefore $E(B-V)_{\mathrm{Total}}=0.30$~$\pm$ 0.08~$\mathrm{mag}$.         

\section{Photometry}
\begin{figure}
\centering
\includegraphics[width=9.0cm]{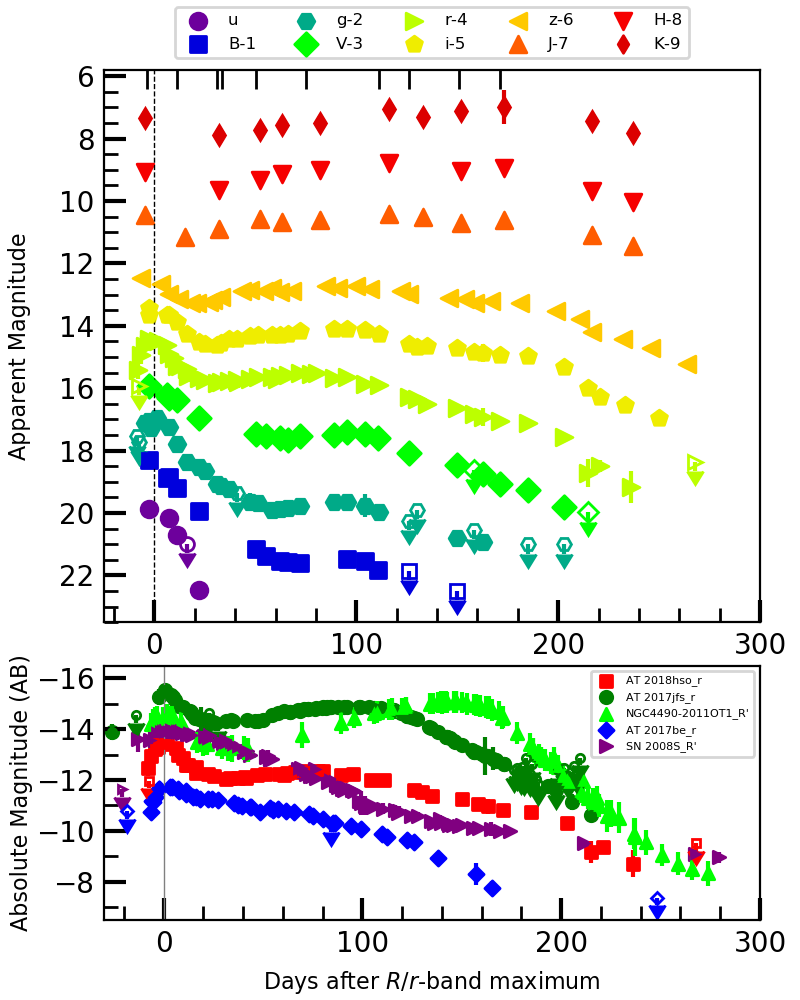}
\caption{Top: $BVugrizJHK$ light curves of AT 2018hso. The epochs of optical spectra are also marked. Bottom:  Sloan $r$-band absolute light curves of AT 2018hso, two ILRTs,  SN~2008S \citep{Botticella2009MNRAS.398.1041B} and AT 2017be \citep{Cai2018MNRAS.480.3424C}, and two LRNe NGC4490-2011OT1 \citep{Pastorello2019arXiv190600812P} and AT 2017jfs \citep{Pastorello2019A&A...625L...8P}. }
\label{lightcurve}
\end{figure}

We started the monitoring of AT~2018hso soon after its discovery, with the follow-up campaign lasting $\sim$ 300 d. 
The photometric data were reduced using the {\sl SNoOpy}\footnote{SNOoPy is a package developed by E. Cappellaro which performs photometry of point-like sources in complex environments using PSF-fitting and template subtraction methods. A package description can be found at \url{http://sngroup.oapd.inaf.it/snoopy.html.}} pipeline, following an ordinary PSF-fitting method as detailed in \citet[][]{Cai2018MNRAS.480.3424C}. The resulting optical and NIR apparent magnitudes are reported in the Appendix (Tables \ref{AT2018hso_opt_LC} and \ref{AT2018hso_nir_LC}). The multi-band light curves are shown in the top panel of Fig. \ref{lightcurve}. The $r$-band light curve rises to the first maximum  (on MJD= 58431.0 $\pm$ 1.0) in $\sim$ 8 d. The peak magnitude is $r \sim 18.40$ mag ($M_{r} \sim -13.93$ mag). After that, the r-band light curve first declines ($\sim 5.3~\mathrm{mag/100d}$), then it slowly rises again to a second, fainter maximum at around 80 d (on MJD $\approx$ 58515).  The magnitude of the second peak is $r \sim 19.57$ mag ($M_{r} \sim -12.76$ mag). A monotonic decline follows, which is initially slow ($\sim1.7~\mathrm{mag/100d}$), but later it rapidly increases to about 3.8~$\mathrm{mag/100d}$. 
While the $i$ and $z$ bands have a similar evolution as the $r$ band, the bluer bands ($B, g,$ and $V$) show a sort of plateau. The few $u$-band detections show a roughly linear decline of $\sim 15.7~\mathrm{mag/100d}$. The NIR light curves resemble those in the red bands, showing a more evident, long-duration ($\sim$ 150 d) second maximum, followed by a moderate decline ($\sim 1.3~\mathrm{mag/100d}$) starting from $\sim$ 190 d. 

We  compare the $r$-band absolute light curve of AT~2018hso with some well-followed gap transients, two ILRTs (SN~2008S and AT~2017be) and two LRNe (NGC4490-2011OT1 and AT 2017jfs) in Fig. \ref{lightcurve} (bottom panel). Their maximum absolute magnitudes range from -12 to -15.5 mag. ILRTs show a single-peaked SN-like light curve, whilst LRNe have double-peaked light curves. We note that AT~2018hso reveals a transitional light curve, between ILRTs and LRNe, with a first sharp blue peak followed by a much shallower and broader red peak.                 


\section{Spectroscopy}\label{spectroscopy}

\begin{figure}
\centering
\includegraphics[width=8.5cm]{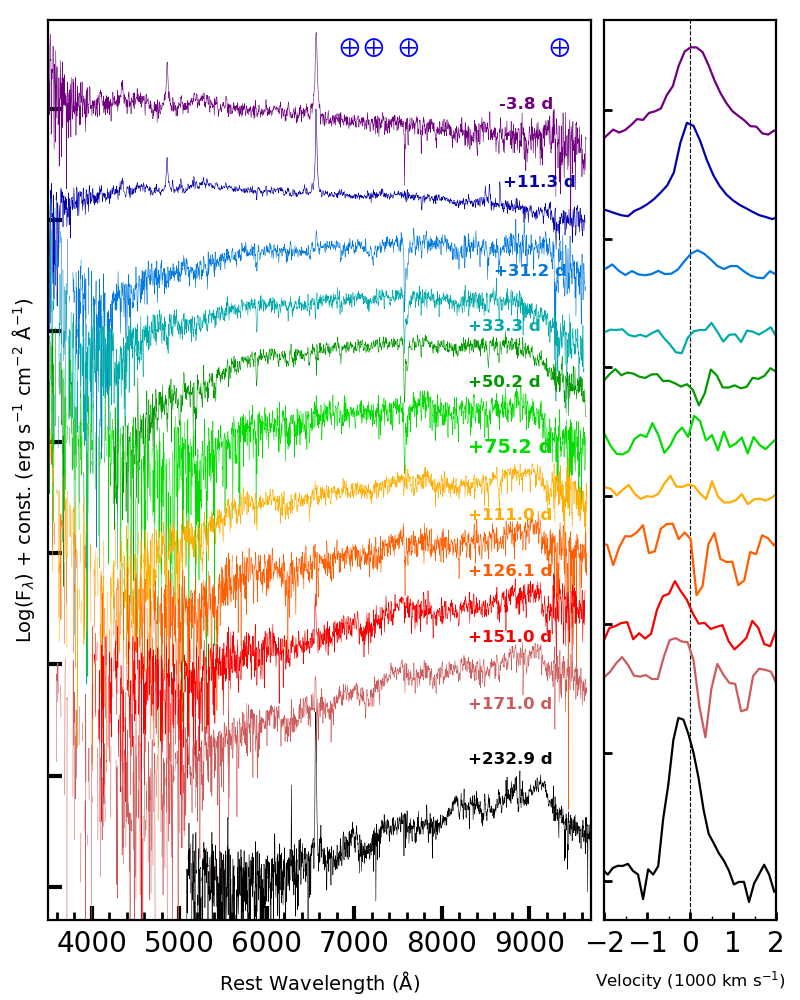}
\caption{Left: Spectral evolution of AT 2018hso, from $-3.8$ to $+232.9$ d. Right: Evolution of the \halpha~profile in the velocity space. The vertical dotted line marks the rest velocity. All spectra are redshift-corrected ($z=0.003536$, see NED). The phases are from the $r$-band maximum ($MJD= 58431.0\pm1.0$).} 
\label{spectseq}
\end{figure}

\begin{figure}
\centering
\includegraphics[width=8.5cm]{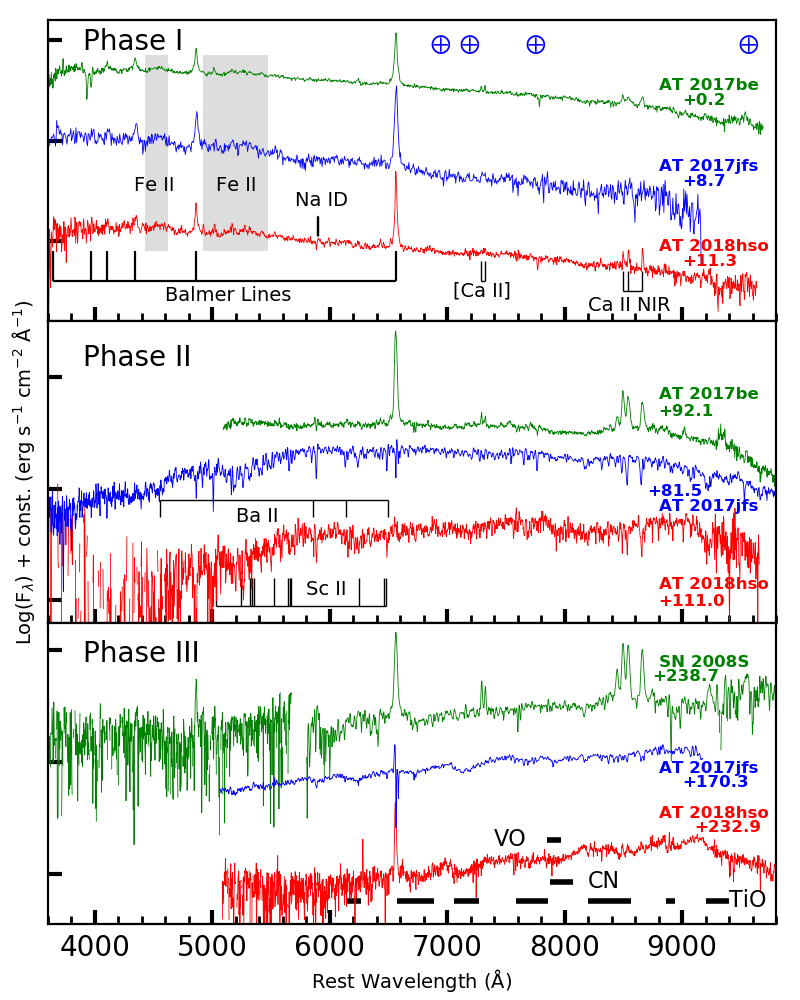}
\caption{Spectral comparison of AT~2018hso with LRN AT~2017jfs, and ILRTs AT~2017be and SN~2008S, with  line identification. The spectra are selected at three representative epochs: Phase I (0 to +10 d; top panel); Phase II ($95\pm15$ d; middle panel); Phase III ($>$170 d; bottom panel). Phases are from the $R/r$-band maximum. All spectra are redshift and reddening corrected. The data of comparison objects are  from \cite{Cai2018MNRAS.480.3424C, Botticella2009MNRAS.398.1041B, Pastorello2019A&A...625L...8P}} 
\label{spectcomp}
\end{figure}

Our spectroscopic campaign spans a period of eight months, from $\sim-4$ to $+233$ d, of the AT~2018hso evolution. We collected 10 epochs of spectroscopic observations with the 2.56-m Nordic Optical Telescope (NOT \footnote{\url{http://www.not.iac.es/}}) equipped with ALFOSC. In addition, through approved Director's Discretionary Time (DDT \footnote{\url{http://vivaldi.ll.iac.es/OOCC/night-cat/ddt-at-gtc/}}, PI: A. Morales-Garoffolo), a very late spectrum was obtained on 2019 June 29 with the 10.4-m Gran Telescopio Canarias (GTC \footnote{\url{http://www.gtc.iac.es/}}) plus OSIRIS (see details in Table \ref{2018hsospecinfo}). The spectra were processed following  standard tasks in {\sc iraf} \footnote{{\sc iraf} is written and supported by the National Optical Astronomy Observatories (NOAO) in Tucson, Arizona. NOAO is operated by the Association of Universities for Research in Astronomy (AURA), Inc. under cooperative agreement with the National Science Foundation.}. The  spectral evolution of AT~2018hso is presented in Fig. \ref{spectseq}, while the comparison with ILRTs AT~2017be and SN~2008S, and LRN AT~2017jfs is shown in  Fig. \ref{spectcomp}. Prominent spectral lines are marked in Fig. \ref{spectcomp}.   
  
The spectral evolution of AT~2018hso is characterised by three distinct phases, in analogy with the behaviour of other LRNe \citep{Pastorello2019arXiv190600812P}. At early epochs (until $\sim$30 d), the spectra show a blue continuum, with prominent Balmer lines, along with a number of Fe {\sc ii} features, Na~{\sc i} D ($\lambda$=5890, 5896 $\AA$), and Ca {\sc ii} emission lines (see Fig. \ref{spectseq} and Fig. \ref{spectcomp},  top panel). The temperature inferred from the spectral continuum, assuming a black-body spectral energy distribution (SED), decreases from 8800 $\pm$ 800 K (at -3.8 d) to 6800 $\pm$ 600 K (at +11.3 d). We measured the full width at half-maximum (FWHM) velocity of H$\alpha$ through a Lorentzian fit, and obtained $v_{\mathrm{FWHM}}$ $\sim$ 500 km s$^{-1}$ (accounting for the instrumental resolution, see Table \ref{2018hsospecinfo}). This early-time velocity is similar to those measured for LRNe, but also some ILRTs, such as AT~2017be \citep[also fitted by Lorentzian function, $\sim$ 500-800 km s$^{-1}$; ][]{Cai2018MNRAS.480.3424C}. A prominent Ca~{\sc ii} NIR triplet ($\lambda$=8498, 8542, 8662 $\AA$) is detected in emission in the AT~2018hso spectra, along with a very weak [Ca {\sc ii}] doublet ($\lambda$=7291, 7328 $\AA$). We remark that a prominent [Ca {\sc ii}]  is an ubiquitous feature in the spectra of ILRTs at all phases \citep{Cai2018MNRAS.480.3424C}. In contrast, it has never been securely identified in LRN spectra so far. Ca~{\sc ii} H\&K ($\lambda$=3934, 3968 $\AA$) is not clearly detected, but the spectra are very noisy below 4000$\AA$. All of this makes  early spectra of AT~2018hso similar to ILRTs (Fig. \ref{spectcomp}, top panel).    

During the rise to the second peak (from $\sim$30 d to 120 d), the spectra experience a dramatic evolution. The continuum becomes redder with the continuum temperature declining from $\sim$ 4500 to 3000 K. Thus, the spectra resemble those of cool stars (K to early M types). H$\alpha$ is marginally detected at +31.2 d, and the emission component disappears until +111.0 d. As a remarkable note, the [Ca {\sc ii}] doublet is no longer detected. In contrast with early spectra, the NIR Ca~{\sc ii} triplet is now seen in absorption. The Ca {\sc ii} lines are formed by radiative de-excitation from upper to lower levels \citep[e.g., Ca {\sc ii} triplet, the transition from $4p ^{2}P_{1/2,3/2}$ to $3d^{2}D_{3/2,5/2}$ levels; ][]{Mallik1997A&AS..124..359M}, in particular [Ca {\sc ii}] doublet originates in a very low density gas with a critical density of about 10$^{7}$~cm$^{-3}$ \citep{Ferland1989ApJ...347..656F}. The major change in those lines indicate the gas densities evolving with time. From the highest SNR spectra, we identify metal lines (Fe~{\sc ii}, Ba~{\sc ii} and Sc~{\sc ii}) in absorption. At this phase, the spectra of AT~2018hso are reminiscent of LRNe, while are clearly different from those observed in ILRTs (see Fig. \ref{spectcomp}, middle panel).         

At very late times (over four months after peak), the spectra of AT 2018hso become even redder ($T$=2050 $\pm$ 200 K at +232.9 d)) and resemble those of a late M star. H$\alpha$ becomes prominent again, with $v_{\mathrm{FWHM}}$ $\sim$ 370 km s$^{-1}$ at +232.9 d as obtained by fitting a Gaussian function. In addition, the H$\alpha$ emission peak appears blue-shifted by about 300 to 400 km s$^{-1}$ at $\sim$150 to 230 d, respectively \footnote{Narrow absorptions are found at  +126.1 d and +171.0 d, are likely due to over-subtraction of contaminating background.} (see right panel of Fig. \ref{spectseq}). This is reminiscent of very late spectra of AT~2017jfs and NGC4490-2011OT1, supporting a LRN classification for AT~2018hso. The asymmetries in the H$\alpha$ profile suggest aspherical geometry of the emitting region, or new dust formation which is obscuring the receding material, or both scenarios \citep{Smith2016MNRAS.458..950S, Pastorello2019arXiv190600812P}. In addition, very late spectra of AT 2018hso show broad molecular bands, such as TiO and possibly VO, CN (Fig. \ref{spectcomp}, bottom panel). This is a key feature that allows us to distinguish LRNe from ILRTs \citep[][]{Pastorello2019arXiv190600812P}. Following the interpretation of \citet{Kaminsk2009ApJS..182...33K}, we may speculate that the TiO molecular bands of AT~2018hso originate in the warm photosphere ($T>2000$ K), but also in the cold outflowing material ($T\sim500$ K) with a velocity about $-$3-400 km s$^{-1}$, where plausibly also VO features form.


\section{Spectral Energy Distribution}\label{sed}

We investigated the SED evolution of AT~2018hso using  photometric data. The SEDs are selected at a few representative epochs along the whole monitoring period. The SEDs are fitted with a single Plankian function, assuming the photosphere radiates as a blackbody. At  early phases, before the blue peak, the object was only occasionally observed. When individual broad-band data are not available, the missing point is estimated by interpolating the data at two consecutive epochs, or by extrapolating from the last available observation. All of this implies larger errors. At 30-110 d, the optical plus NIR SEDs are well reproduced with a single blackbody. Instead, although at later phases blue-band ($u,B,g,V$) data are not available, a single blackbody does not appear to accurately fit the observed SEDs. This is even more evident during the final steep lightcurve decline (at $\sim$210 d). This is suggestive of a second blackbody component peaking at much longer wavelengths (mid to far IR), although it cannot be confirmed due to the incomplete photometric coverage.   

The temporal evolution of blackbody temperature is shown in Fig. \ref{sed} (top panel). The temperature rapidly declines from $\sim$8000~K at maximum to $\sim$4000~K at about 30 d. Then, it declines more slowly to $\sim$3000~K until 180 d. A steeper temperature drop (to around 2500~K) follows at $\sim$210~d. The inferred bolometric luminosity, obtained integrating the SEDs along the full electromagnetic spectrum, is shown in the middle panel of Fig. \ref{sed}. The luminosity at peak is about $10^{41}$ erg s$^{-1}$, then it rapidly declines to a minimum ($L\sim 3.2 \times 10^{40}$ erg s$^{-1}$), rises again to the second, red maximum ($L \sim4.4\times10^{40}$ erg s$^{-1}$), and finally rapidly drop to $L \sim 1.6 \times 10^{40}$ erg s$^{-1}$. The inferred radius, obtained  through the Stefan-Boltzmann law ($L=4\pi R^2 \sigma T^4$; where $\sigma$ is the Stefan-Boltzmann constant), is shown in the bottom panel of Fig. \ref{sed}. The radius remains roughly constant at $\sim2\times10^{14}$ cm (almost 2\,900~R$_{\odot}$) during the first $\sim$30 d, then it rapidly increases to $\sim 5.3 \times 10^{14}$ cm (exceeding 7\,600~R$_{\odot}$) at $\sim$80 d, to increase by a modest amount later. A significant increase in the radius is observed after $\sim$ 210 d (although affected by large uncertainties), and exceeds $\sim 7.2 \times 10^{14}$ cm ($R$ $\sim10\,350$~R$_{\odot}$).        

The above parameters of AT 2018hso are compared with those of a few ILRTs and LRNe in Fig. \ref{sed}. While the temperature evolution is not relevant to discriminate the two classes, the bolometric light curve and the radius evolution are markedly different in LRNe and ILRTs.
Specifically, ILRT light curves decline monotonically after maximum, while LRN ones reveal a second bolometric peak. The difference between LRNe and ILRTs is even more evident in the radius evolution. While in ILRTs the radius at the photosphere of the hot blackbody component declines by a factor of two \citep[see ][]{Botticella2009MNRAS.398.1041B}, it increases by one order of magnitude during the LRN evolution, with a major increase during the rise to the second photometric peak. A further increase in the value of the radius is observed at very late phases, during the final steep optical decline. 
This is observed also in AT~2017jfs, at the time of a NIR brightening. In this phase, the H$\alpha$ profile is blue-shifted and the spectra show molecular bands. The different light curves and radius evolution in ILRTs and LRNe suggest that the two classes of gap transients are regulated by different physical mechanisms. The contraction of ILRT radius after peak is likely a consequence of the outer layers becoming optically thin with the expansion, with the photosphere receding into deeper ejecta. The more complex evolution of the LRN radii, already noted by \citet{Blagorodnova2017ApJ...834..107B} for M101-2015OT1, is more difficult to explain. In particular, while M31-2015LRN \citep{MacLeod2017ApJ...835..282M} showed a contraction of the radius at late phases consistent with the photosphere receding with expansion, brighter LRNe are characterised by at least two phases with an almost constant radius, occurring soon after the first and the second peaks. After the first peak, the expanding ejecta lower their opacity and thus the radius in mass coordinate decreases. When these ejecta reach the material expelled during the CE phase, this gas is initially heated, and a second photosphere forms far out leading the radius to increase. The ejecta plus CE shocked region expands remaining optically thick, hence the photospheric radius grows.     

\begin{figure}
\centering
\includegraphics[width=8.7cm]{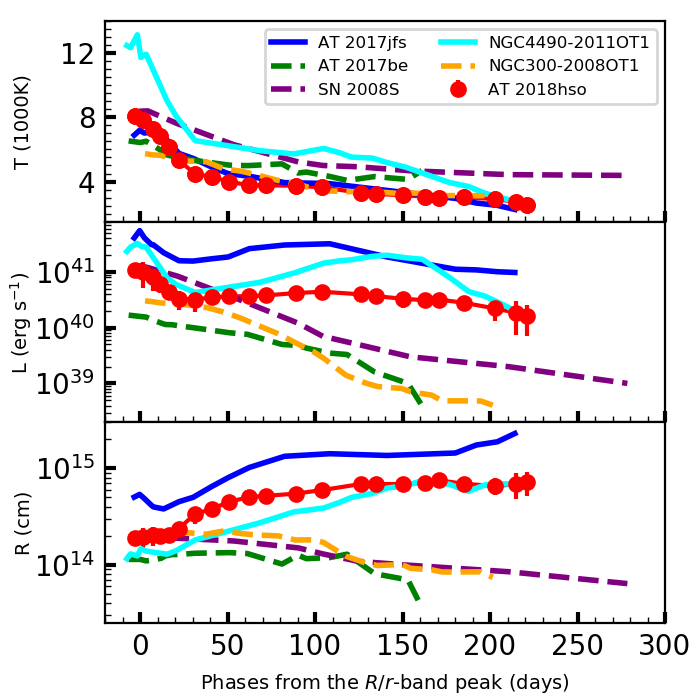}
\caption{Comparison of \object{AT 2018hso} with three ILRTs (\object{SN 2008S},  \object{NGC300-2008OT1}, \object{AT 2017be}; dashed lines), and two LRNe (\object{AT~2017jfs}, \object{NGC4490-2011OT1}; solid lines).
Top: Blackbody temperature evolution	. Middle: Bolometric light curves. Bottom: Evolution of  the photospheric radii. The data are  from \citet{Botticella2009MNRAS.398.1041B}, \citet{Humphreys2011ApJ...743..118H}, \citet{Cai2018MNRAS.480.3424C}, and \citet{Pastorello2019A&A...625L...8P, Pastorello2019arXiv190600812P}. } 
\label{sed}
\end{figure}

\section{On the nature of AT~2018hso}\label{discussion}
AT~2018hso shows transitional light curves between ILRTs and LRNe, and  early spectra similar to those of ILRTs. However, we favour a LRN classification for the following reasons:  
\begin{itemize}
\item The spectra experience a major temporal evolution. Early-time spectra resemble those of ILRTs. Nonetheless, the [Ca {\sc ii}] doublet vanishing with time, the tremendous change of H$\alpha$ profile, and the late appearance of molecular bands support the LRN scenario for AT~2018hso.  
\item In analogy to other LRNe, it peaks at an absolute magnitude of $M_{\mathrm{r}}$ $\sim-13.93$ mag after a fast rise lasting about 8 d.   
\item The SED and the evolution of the photospheric radius are similar to those observed in other LRNe. 
\item The decay time from peak luminosity ($L_{\mathrm{peak}}$$\approx$$10^{41}$ erg s$^{-1}$) to 0.5$L_{\mathrm{peak}}$ is around 15 d, located near the LRN region in the luminosity vs. $\tau_{\mathrm{0.5dex}}$ diagram of \citet[][see their figure 1]{Pastorello2019NatAs...3..676P}. 
\item  The outflow velocity of AT 2018hso is around 400 km s$^{-1}$, placing it  slightly below NGC4490-2011OT1 in the $L_{\mathrm{peak}}$ vs. $v_{\mathrm{out}}$ diagram of \citet{Mauerhan2018MNRAS.473.3765M}, but still well aligned with other LRNe.
\end{itemize}

To add further support to the LRN classification, we inspected HST images obtained from the Hubble Legacy Archive. NGC3729 was observed on 2004 Nov. 17 with HST+ACS. Unfortunately, these observations were quite shallow, consisting of 2$\times$350s in F658N, and 120s in F814W. In order to locate the position of the SN on these images, we aligned the drizzled F814W HST image to the NOT+AFOSC $i$-band image taken on 2019 Jan 14. Only five sources were used for the alignment, which has an RMS scatter of 0.12\arcsec. In light of the small number of fiducial sources for the alignment, we caution that there could be a larger systematic uncertainty in position. A bright source is detected with S/N $\approx$17, within 2$\sigma$ of the transformed SN position (see Fig. \ref{18hsoProgenitor} in appendix). Using {\sc dolphot}, we measure  $F658N=23.29\pm0.16$~mag and $F814W = 23.56\pm0.07$~mag  (Vegamag, in the HST system), for the object. At the adopted distance, twe obtain an absolute magnitude of $I \sim -8$. This is  at least plausibly consistent with a pre-merger system that has started to brighten around fifteen years before the coalescence, similar to what observed in the 2015 LRN in M101 \citep[][]{Blagorodnova2017ApJ...834..107B}. However, in the absence of more precise astrometry, and at the spatial resolution of our data at the distance of NGC 3729, we cannot exclude that this is a cluster, or an unrelated source. Late time observations after the merger has faded and cooled will reveal if this was the progenitor system, as we would expect this source to become dust-enshrouded, disappearing (or at least fading) in the optical. 
\citet{Pastorello2019arXiv190600812P} studied a large sample of LRNe, supporting for them a stellar merger scenario in a binary system  following the ejection of a CE \citep[e.g.][]{Tylenda2011A&A...528A.114T,Kochanek2014MNRAS.443.1319K,Pejcha2016MNRAS.461.2527P,Pejcha2017ApJ...850...59P,MacLeod2017ApJ...835..282M}.  \citet{Kochanek2014MNRAS.443.1319K} and \citet{Smith2016MNRAS.458..950S} suggested that energetic LRN events, such as NGC4490-2011OT1, are likely the outcome of massive (10 $\leq M \leq$ 50 M$_{\odot}$) binary mergers. In this context, as AT~2018hso has a plausible luminous source detection and only marginally fainter than NGC4490-2011OT1, its progenitor system was very likely quite massive \citep{Kochanek2014MNRAS.443.1319K} .

\begin{figure}
\centering
\includegraphics[width=8.7cm]{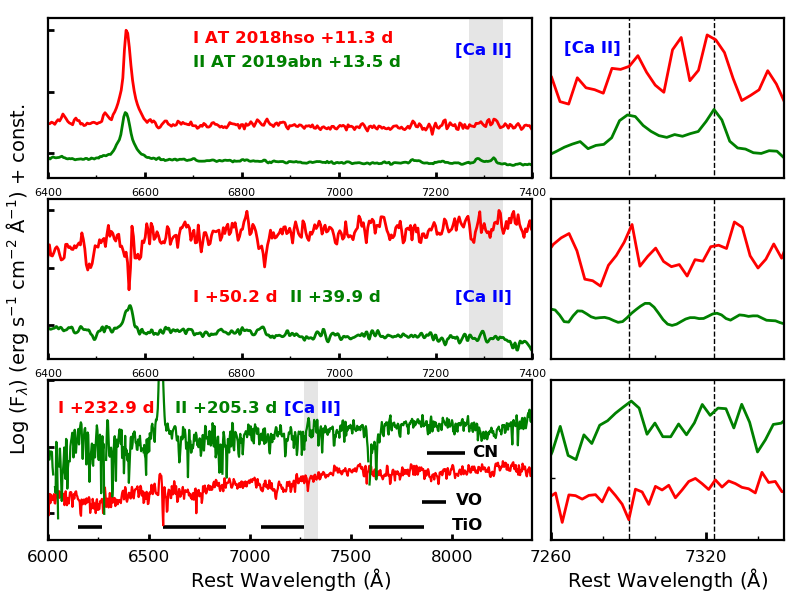}
\caption{Spectral comparison of \object{AT 2018hso} and AT~2019abn. Left: spectral evolution at three representative phases:  $\sim$12$\pm$1 d (top);  $\sim$45$\pm$5 d (middle); $\geq$200 d (bottom). The molecular bands are identified in the left bottom panel. The [Ca {\sc ii}] doublet is marked with a shaded grey region in three left panels. Right: blow-up of the [Ca {\sc ii}] doublet profile at the the same phases. Phases are from their $r$-band maximum. } 
\label{HsoAbn}
\end{figure}

From the study of the spectro-photometric evolution of AT~2018hso and the possible detection of the progenitor system in archive {\sc hst} images, we provided reasonable arguments for classifying AT~2018hso as a LRN, although with somewhat peculiar characteristics. So far, to our knowledge, only two other objects have controversial ILRT/LRN classifications. Although M85-2006OT1 was proposed to be a LRN \citep{Kulkarni2007Natur.447..458K,Rau2007ApJ...659.1536R}, \citet{Pastorello2007Natur.449E...1P} questioned this classification, and an ILRT scenario is still plausible from observational arguments \citep{Kochanek2014MNRAS.443.1319K,Pastorello2019arXiv190600812P}. Another gap transient, AT~2019abn, was classified as an ILRT by \citet{Jencson2019ApJ...880L..20J} on the basis of follow-up observations lasting $\sim$ 110 d and a detailed study of the progenitor in pre-outburst stage. 
However, the initial weakness of the [Ca {\sc ii}] doublet and its later disappearance (see Fig. \ref{HsoAbn}) are reminiscent of LRNe. Only the lack of strong TiO and VO bands in the optical spectra will provide further support to the ILRT classification. All of this suggests the existence of a grey zone in our ability to distinguish the different classes of gap transients.

We have limited knowledge on ILRTs and LRNe due to the still modest number of discoveries and incomplete data sets. Well-sampled light curves in a wide range of wavelengths, high-resolution spectroscopy, and detailed modelling are crucial to clarify the nature of both families of gap transients. The support of future generation instruments such as the Large Synoptic Survey Telescope \citep[][]{LSST2009arXiv0912.0201L} and the Wide Field Infrared Survey Telescope \citep[WFIRST; ][]{Spergel2015arXiv150303757S} is essential to  increase the number of well-monitored gap transients, crucial to fine tune existing theoretical models.

\begin{acknowledgements}
We thank K. Maguire, X-F. Wang for useful discussions. Y-Z.C is supported by the China Scholarship Council (No. 201606040170). KM and SJP are supported by H2020 ERC grant no.~758638. MF is supported by a Royal Society - Science Foundation Ireland University Research Fellowship. SB and AF are partially supported by PRIN-INAF 2017 of Toward the SKA and CTA era: discovery, localization, and physics of transient sources.(PI: M. Giroletti). T.R. acknowledges the financial support of the Jenny and Antti Wihuri and the Vilho, Yrj{\"o} and Kalle V{\"a}is{\"a}l{\"a} foundations. AR acknowledge financial support by the "Millennium Institute of Astrophysics (MAS)" of the Iniciativa Cient\`{i}fica Milenio. GC acknowledges support from European Research Council Consolidator Grant 647208. MG is supported by the Polish NCN MAESTRO grant 2014/14/A/ST9/00121. FO acknowledge the support of the H2020 Hemera program, grant agreement No 730970. The Nordic Optical Telescope (NOT), operated by the NOT Scientific Association at the Spanish Observatorio del Roque de los Muchachos of the Instituto de Astrofisica de Canarias. Observations from the NOT were obtained through the NUTS1 and NUTS2 collaboration which are supported in part by the Instrument Centre for Danish Astrophysics (IDA). The data presented here were obtained [in part] with ALFOSC, which is provided by the Instituto de Astrofisica de Andalucia (IAA) under a joint agreement with the University of Copenhagen and NOTSA. The Liverpool Telescope is operated on the island of La Palma by Liverpool John Moores University in the Spanish Observatorio del Roque de los Muchachos of the Instituto de Astrofisica de Canarias with financial support from the UK Science and Technology Facilities Council. Based on observations made with the GTC telescope in the Spanish Observatorio del Roque de los Muchachos of the Instituto de Astrofssica de Canarias, under Director's Discretionary Time (GTC2019-127; PI: A. Morales-Garoffolo). This research makes use of Lasair data (https://lasair.roe.ac.uk/), which is supported by the UKRI Science and Technology Facilities Council and is a collaboration between the University of Edinburgh (grant ST/N002512/1) and Queen's University Belfast (grant ST/N002520/1) within the LSST:UK Science Consortium.  
 \end{acknowledgements}


\bibliographystyle{aa}
\bibliography{paperLRNe}

\begin{appendix}
\section{Additional Data}
\clearpage

\begin{figure}
\centering
\includegraphics[width=8.9cm]{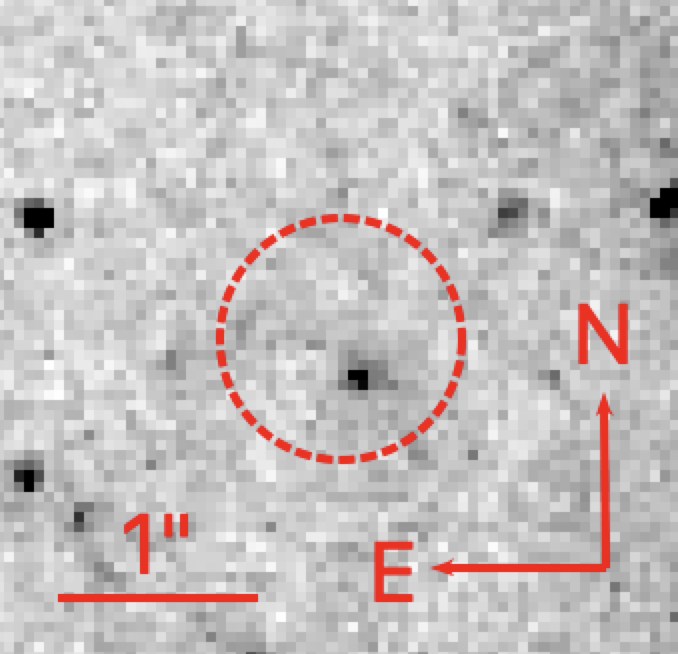}
\caption{Pre-outburst archive image of AT 2018hso obtained with HST+F814W. The red dashed circle shows 5x RMS (0.6\arcsec) uncertainty in the position.} 
\label{18hsoProgenitor}
\end{figure}   

\begin{table*}
\caption{General information of the spectroscopic observations of AT~2018hso.}
\label{2018hsospecinfo}
\begin{tabular}{@{}cccccccc@{}}
\hline
Date &   & Phase $^a$  & Telescope+Instrument & Grism+Slit  & Spectral range & Resolution & Exp. time   \\ 
        &          & (days) &                               &                            & ($\AA$)                  & ($\AA$)           & (s)                \\ 
\hline
20181105 &  58427.2 & $-3.8$  &   NOT+ALFOSC     & gm4+1.0"            & 3400-9650     &  15  & 1670  \\
20181120 &  58442.3 & +11.3   &    NOT+ALFOSC    & gm4+1.0"            &  3500-9630    &  15 & 3600 \\
20181210 &  58462.2 & +31.2  &    NOT+ALFOSC     & gm4+1.0"            & 3400-9600    &  15  &  3600 \\
20181212 &  58464.3 & +33.3  &    NOT+ALFOSC     & gm4+1.3"            & 3400-9600    &  18  &  3600\\
20181229 &  58481.2 & +50.2  &    NOT+ALFOSC     & gm4+1.0"            & 3400-9600    &  15  &  3600\\
20190123 &  58506.2 & +75.2  &    NOT+ALFOSC     & gm4+1.0"            &  3400-9650   &  15  &  3600 \\
20190228 &  58542.0 & +111.0 &    NOT+ALFOSC     & gm4+1.0"            & 3400-9650    &  15  &  3600\\
20190315 &  58557.1 & +126.1 &    NOT+ALFOSC     & gm4+1.0"           & 3400-9650    &  15  &  3600\\
20190409 &  58582.0 & +151.0 &    NOT+ALFOSC     & gm4+1.0"           & 4000-9600    &  15  &  3600\\
20190428 &  58602.0 & +171.0 &    NOT+ALFOSC     & gm4+1.0"           &  3600-9600   &  15  &  3600 \\
20190629 &  58663.9 & +232.9 &    GTC+OSIRIS       & R1000R+1.0"     &  5100-10350 &   8   &  3$\times$1800 \\
\hline
\end{tabular}

\medskip

$^a$ Phases are relative to $r$-band maximum (MJD$_{\mathrm{AT 2018hso}}$= 58431.0 $\pm$ 1.0).\\ 
\end{table*}

\begin{table*}
\begin{minipage}{175mm}                                  
\caption{Optical ($BVugriz$) light curves of AT~2018hso.}
\label{AT2018hso_opt_LC}  
\scalebox{0.8}{
\begin{tabular}{@{}cccccccccccl@{}}
\hline
Date         &     MJD      &  $B$(err) &  $V$(err)         & $u$(err) &  $g$(err)         &  $r$(err)        &  $i$(err)   &  $z$(err)       & Instrument \\
20181031& 58422.487&  --            & --             &  --             &  $>$19.5      &   --            & --            & --              &  1           \\
20181031& 58423.036&  --            & --             &  --             &   --          &  19.410 (0.3  ) & --            & --              & 1 	      \\
20181101& 58423.466&  --            & --             &  --             &  $>$19.7      &    --           & --            & --              &1 	      \\
20181101& 58423.534&  --            & --             &  --             &   --          &   $>$19.9       & --            & --              &1 	      \\
20181102& 58424.529&  --            & --             &  --             &   --          &  18.920 (0.3  ) & --            & --              & 1 	      \\
20181104& 58426.487&  --            & --             &  --             & 19.103 (0.096)&    --           & --            & --              & 1 	      \\
20181104& 58426.530&  --            & --             &  --             &   --          &  18.656 (0.081) & --            & --              & 1 	      \\
20181106& 58428.270& 19.310 (0.047) & 18.926 (0.040) &  19.876 (0.119) & 19.088 (0.042)&  18.627 (0.035) & 18.638 (0.044)&  18.508 (0.053) & 2\\
20181107& 58429.469&  --            & --             & --              & 19.259 (0.115)&   --            & --            & --              & 1 	      \\
20181110& 58432.456&  --            & --             & --              & 18.960 (0.088)&   --            & --            & --              & 1 	      \\
20181110& 58432.547&  --            & --             & --              &   --          &  18.483 (0.066) & --            & --              & 1 	      \\
20181115& 58437.210& 19.864 (0.030) & 19.206 (0.039) & --              &   --          &  18.612 (0.114) & 18.666 (0.054)& --              & 3   \\
20181116& 58438.210& 19.838 (0.154) & 19.321 (0.070) &  20.147 (0.157) & 19.236 (0.050)&  18.895 (0.037) & 18.665 (0.039)&  18.671 (0.056) & 2          \\
20181118& 58440.530&    --          &    --          &     --          &   --          &  19.025 (0.078) &   --          &    --           & 1 	      \\
20181120& 58442.215& 20.212 (0.088) & 19.378 (0.081) &  20.709 (0.089) & 19.777 (0.031)&  19.285 (0.040) & 18.870 (0.059)&  18.971 (0.171) & 3   \\
20181121& 58443.431&   --           &   --           &     --          &    --         &  19.291 (0.168) &   --          &   --            & 1 	      \\
20181125& 58447.270&   --           &   --           &   $>$21.0       & 20.352 (0.120)&  19.378 (0.058) & 19.262 (0.091)&  19.130 (0.116) & 2 	      \\
20181125& 58447.470&   --           &   --           &     --          &   --          &  19.620 (0.186) &   --          &     --          & 1 	      \\
20181201& 58453.240& 20.947( 0.105) & 19.955 (0.109) &  22.482 (0.148) & 20.509 (0.069)&  19.659 (0.077) & 19.529 (0.057)&  19.254 (0.042) & 3   \\
20181204& 58456.175&    --          &   --           &    --           & 20.638 (0.101)&  19.735 (0.059) & 19.582 (0.113)&  19.276 (0.058) & 2 	      \\
20181210& 58462.190&    --          &   --           &    --           & 21.082 (0.089)&  19.844 (0.043) & 19.627 (0.026)&  19.248 (0.032) & 3   \\
20181212& 58464.180&    --          &   --           &    --           & 21.126 (0.132)&  19.807 (0.062) & 19.546 (0.071)&  19.164 (0.093) & 2 	      \\
20181216& 58468.175&    --          &   --           &    --           & 21.233 (0.111)&  19.783 (0.057) & 19.423 (0.054)&  19.068 (0.079) & 2 	      \\
20181220& 58472.110&    --          &   --           &    --           &  $>$21.3      &  19.746 (0.062) & 19.407 (0.044)&     --          & 2 	      \\
20181220& 58472.550&   --          &   --            &     --          &   --                   &  19.811 (0.194) &   --                  &     --                   & 1 	      \\
20181226& 58478.135&    --          &   --           &    --           & 21.637 (0.267)&  19.702 (0.073) & 19.337 (0.053)&  18.883 (0.131) & 2          \\
20181229& 58481.135&  22.168 (0.229)& 20.475 (0.099) &    --           &    --         &     --          &    --         &     --          & 3   \\
20181230& 58482.115&    --          &    --          &    --           & 21.690 (0.120)&  19.649 (0.075) & 19.307 (0.032)&  18.861 (0.086) & 2       \\
20190103& 58486.225&  22.381 (0.171)& 20.528 (0.048) &    --           &    --         &     --          &    --         &     --          & 3   \\
20190106& 58489.085&    --          &    --          &    --           & 21.908 (0.180)&  19.644 (0.107) & 19.308 (0.042)&  18.878 (0.070) & 2       \\
20190108& 58491.557&   --                    &   --                    &     --          &   --                   &  19.698 (0.130) &   --                   &     --                   & 1 	      \\
20190110& 58493.235&  22.544 (0.142)& 20.608 (0.045) &    --           & 21.861 (0.100)&  19.607 (0.079) & 19.284 (0.063)&  18.772 (0.048) & 3   \\
20190114& 58497.175&  22.559 (0.134)& 20.644 (0.085) &    --           & 21.843 (0.084)&  19.567 (0.113) & 19.268 (0.113)&  18.926 (0.067) & 3   \\
20190120& 58503.175&  22.594 (0.195)& 20.543 (0.040) &    --           & 21.788 (0.154)&  19.510 (0.110) & 19.156 (0.054)&  18.875 (0.034) & 3   \\
20190125& 58508.484&   --                    &   --                    &     --          &   --                   &  19.543 (0.178) &   --                   &     --                   & 1 	      \\
20190128& 58511.386&   --                    &   --                    &     --          &   --                   &  19.516 (0.122) &   --                   &     --                   & 1 	      \\
20190206& 58520.145&    --          & 20.494 (0.043) &    --           & 21.657 (0.074)&  19.672 (0.026) & 19.091 (0.020)&  18.737 (0.040) & 3   \\
20190212& 58526.175&  22.479 (0.197)& 20.392 (0.059) &    --           & 21.637 (0.068)&  19.632 (0.041) & 19.093 (0.013)&  18.773 (0.014) & 3   \\
20190221& 58535.095&  22.529 (0.223)& 20.452 (0.051) &    --           & 21.767 (0.366)&  19.869 (0.051) & 19.146 (0.040)&  18.715 (0.040) & 3   \\
20190227& 58541.995&  22.824 (0.249)& 20.580 (0.049) &    --           & 21.968 (0.115)&  19.900 (0.032) & 19.258 (0.020)&  18.832 (0.038) & 3   \\
20190315& 58557.085&   $>$22.9      & 21.064 (0.124) &    --           & $>$22.2       &  20.288 (0.051) & 19.591 (0.037)&  18.887 (0.048) & 3   \\
20190318& 58560.875&   --           &  --            &    --           & $>$21.9       &  20.358 (0.139) & 19.675 (0.079)&  18.983 (0.073) & 2       \\
20190323& 58565.950&   --           &  --            &    --           &    --         &  20.512 (0.095) & 19.659 (0.051)&     --          & 2       \\
20190407& 58580.995&   --           &  --            &    --           & 22.790 (0.200)&  20.631 (0.138) & 19.712 (0.045)&  19.120 (0.073) & 3   \\
20190408& 58581.005&   $>$23.5      & 21.469 (0.075) &    --           &    --         &    --           &   --          &      --         & 3   \\
20190416& 58589.080&   --           &  $>$21.6       &    --           & $>$22.5       &  20.835 (0.131) & 19.835 (0.043)&  19.133 (0.057) & 3   \\
20190420& 58593.930&   --           & 21.755 (0.204) &    --           & 22.936 (0.221)&  20.918 (0.284) & 19.863 (0.290)&  19.267 (0.059) & 3   \\
20190428& 58601.950&   --           & 22.061 (0.211) &    --           &   --          &  21.060 (0.086) & 19.932 (0.033)&  19.202 (0.036) & 3   \\
20190512& 58615.890&   --           & 22.249 (0.176) &    --           &  $>$23.0      &  21.126 (0.090) & 19.982 (0.036)&  19.279 (0.032) & 3   \\
20190530& 58633.885&   --           & 22.819 (0.319) &    --           &  $>$23.1      &  21.577 (0.119) & 20.326 (0.044)&  19.528 (0.043) & 3   \\
20190611& 58645.925&    --          &  $>$23.0       &    --           &   --          &  22.007 (0.073) & 21.001 (0.054)&  19.750 (0.054) & 3   \\
20190617& 58651.945&    --          &   --           &    --           &   --          &  22.497 (0.183) & 21.269 (0.051)&  20.203 (0.052) & 3   \\
20190629& 58663.880&    --          &   --           &    --           &   --          &    --           & 21.544 (0.118)&     --          & 4  \\
20190702& 58666.895&    --          &   --           &    --           &   --          &  23.172 (0.521) &    --         &  20.433 (0.058) & 3  \\
20190716& 58680.915&    --          &   --           &    --           &   --          &    --           & 21.952 (0.243)&  20.705 (0.091) & 3  \\
20190803& 58698.875&    --          &   --           &    --           &   --          &    --           & $>$22.4       &  21.216 (0.127) & 3  \\
\hline

\hline
\end{tabular}   
}
\medskip

$1$ ZTF data from the Palomar 1.2-m Oschin Telescope equipped with ZTF-Cam, taken from {\sc Lasair} \footnote{{\sc Lasair}: \url{https://lasair.roe.ac.uk/object/ZTF18acbwfza/}} \citep{Smith2019RNAAS...3a..26S} and Transient Name Server ({\sc tns} \footnote{{\sc tns}: \url{https://wis-tns.weizmann.ac.il/object/2018hso}}).\\
$2$ The 2-m fully automatic Liverpool Telescope (LT) equipped with IO:O, located at Roque de los Muchachos Observatory (La Palma,
Canary Islands, Spain). \\
$3$ The 2.56-m Nordic Optical Telescope (NOT) equipped with ALFOSC, located at Roque de los Muchachos Observatory (La Palma,
Canary Islands, Spain). \\
$4$ The 10.4-m Gran Telescopio Canarias (GTC) with OSIRIS, located at Roque de los Muchachos Observatory (La Palma,
Canary Islands, Spain). \\

\end{minipage}
\end{table*} 

\begin{table*}
\begin{minipage}{175mm}
\caption{NIR ($JHK$) light curves of AT~2018hso. }
\label{AT2018hso_nir_LC}
\begin{tabular}{@{}cccccl@{}}
\hline
Date & MJD & $J$(err) & $H$(err) & $K$(err) & Instrument \\
\hline
20181104 &58426.217 & 17.462 (0.034) & 17.074 (0.024) & 16.332 (0.026) &  NOTCAM \\
20181124 &58446.150 & 18.143 (0.091) &  --                     &  --                     &  NOTCAM \\
20181211 &58463.153 & 17.890 (0.104) & 17.641 (0.119) & 16.877 (0.126) &  NOTCAM \\
20181231 &58483.070 & 17.585 (0.117) & 17.329 (0.115) & 16.725 (0.111) &  NOTCAM \\
20190111 &58494.160 & 17.673 (0.098) & 17.131 (0.098) & 16.560 (0.163) &  NOTCAM \\
20190130 &58513.213 & 17.598 (0.097) & 17.014 (0.102) & 16.506 (0.125) &  NOTCAM \\
20190305 &58547.080 & 17.424 (0.092) & 16.782 (0.093) & 16.060 (0.119) &  NOTCAM \\
20190322 &58564.045 & 17.518 (0.122) &  --                     & 16.301 (0.106) &  NOTCAM \\
20190410 &58583.053 & 17.693 (0.035) & 17.053 (0.124) & 16.106 (0.114) &  NOTCAM \\
20190430 &58603.950 & 17.599 (0.074) & 16.933 (0.096) & 15.991 (0.547) &  NOTCAM \\
20190613 &58647.970 & 18.091 (0.122) & 17.666 (0.119) & 16.437 (0.121) &  NOTCAM \\
20190703 &58667.933 & 18.430 (0.098) & 18.030 (0.064) & 16.833 (0.155) &  NOTCAM \\

\hline
\end{tabular}

\medskip
NOTCAM: The 2.56-m Nordic Optical Telescope (NOT) equipped with NOTCAM, located at Roque de los Muchachos Observatory (La Palma,
Canary Islands, Spain).\\ 
\end{minipage}
\end{table*}
 
\end{appendix}

\end{document}